\documentclass{article}

 \usepackage[final]{neurips_2025}

\usepackage[utf8]{inputenc} 
\usepackage[T1]{fontenc}    
\usepackage{hyperref}       
\usepackage{url}            
\usepackage{booktabs}       
\usepackage{amsfonts}       
\usepackage{nicefrac}       
\usepackage{microtype}      
\usepackage{xcolor}         
\usepackage{float}
\usepackage{graphicx}
\usepackage{makecell}
\usepackage{enumitem}
\usepackage{bm}
\usepackage{amsmath}
\usepackage{appendix}
\usepackage{comment}

\title{zkFinGPT: Zero-Knowledge Proofs for Financial Generative Pre-trained Transformers}

\author{%
  Xiao-Yang Liu$^1$\thanks{Corresponding author.}~~, Ningjie Li$^2$, Keyi Wang$^1$, Xiaoli Zhi$^2$, Weiqin Tong$^2$\\
  $^1$SecureFinAI Lab, Columbia University, New York, NY 10027 \\
  $^2$School of Computer Engineering and Science, Shanghai University, Shanghai 201824 \\
  Emails: \{XL2427, KW2914\}@columbia.edu \\
}

\begin{document}

\maketitle

\begin{abstract}

Financial Generative Pre-trained Transformers (FinGPT) with multimodal capabilities are now being increasingly adopted in various financial applications. However, due to the intellectual property of model weights and the copyright of training corpus and benchmarking questions, verifying the legitimacy of GPT's model weights and the credibility of model outputs is a pressing challenge.  In this paper, we introduce a novel zkFinGPT scheme that applies zero-knowledge proofs (ZKPs) to high-value financial use cases, enabling verification while protecting data privacy. We describe how zkFinGPT will be applied to three financial use cases. Our experiments on two existing packages reveal that zkFinGPT introduces substantial computational overhead that hinders its real-world adoption. E.g., for LLama3-8B model, it generates a commitment file of $7.97$MB using $531$ seconds, and takes $620$ seconds to prove and $2.36$ seconds to verify.

\end{abstract}

\section{Introduction}

Financial Generative Pre-trained Transformers (FinGPT) \cite{wu2023bloomberggpt, liu2023fingpt,liu2024fingpt} and agents with multimodal capabilities \cite{yanglet2025multimodal} have been increasingly applied to various financial applications. However, privacy and trustworthiness have raised public concerns. The most prominent issue is the active lawsuits between model producer companies and traditional content publishers \cite{lawsuit2025}. Many leading AI companies, such as OpenAI, Microsoft, and Perplexity.ai, are being sued by publishers for copyright infringement, involving hundreds of billions of dollars. Closed-source models' weights, such as GPT-4 \cite{openai2024gpt4technicalreport}, are intellectual property. Therefore, courts need an effective method to verify GPTs' outputs while ensuring the confidentiality of model weights. In high-stakes domains of finance and healthcare, it requires the credibility of model outputs without compromising data privacy \cite{colin2024leaderboard}.

\begin{figure}[h]
\vspace{-0.1in}
    \centering
    \includegraphics[width=0.9\textwidth]{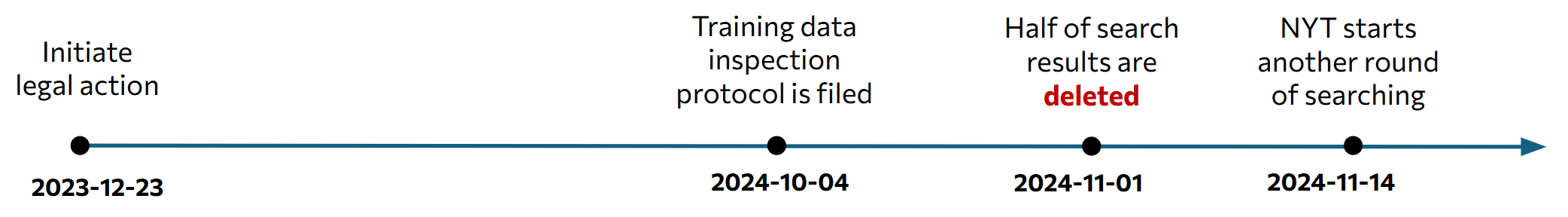}  
    \caption{Timeline of the trial between New York Times and OpenAI.} 
    \label{fig:timeline}
    \vspace{-0.15in}
\end{figure}

In this paper, we introduce a novel zkFinGPT scheme that uses zero-knowledge proof (ZKP) \cite{sun2024zkllm, qu2025zkgpt} to make FinGPT's inference process publicly verifiable. We demonstrate the use of zkFinGPT in three financial cases: 1) the lawsuit between NYT and OpenAI \cite{nyt2023lawsuit}, 2) the credibility of FinGPT's results on copyrighted exam questions \cite{patel2025reasoning}, and 3) the protection of trading strategies in the FinRL Contest \cite{wang2025finrlcontestsbenchmarkingdatadriven}. The zkFinGPT verifies the legitimacy of the model weights and the credibility of its outputs. Our experiments on two existing packages reveal that zkFinGPT introduces substantial computational overhead. E.g., for the LLama3-8B model, it generates a commitment file of $7.97$MB using $531$ seconds, and takes $620$ seconds to prove and $2.36$ seconds to verify.

\section{Financial Use Cases}

\subsection{Use Case I: Copyright Lawsuit of Training Corpus}


There is an ongoing legal battle between model producer companies and traditional content publishers. NYT sued OpenAI and Microsoft together \cite{nyt2023lawsuit}, accusing them of using millions of copyrighted articles to train the GPT-4 model without authorization. During the trial procedure in Fig. \ref{fig:timeline}, the court required OpenAI to set up two servers as a “sandbox” where NYT lawyers examined the training corpus remotely.  However, OpenAI engineers accidentally deleted the operation logs on the servers \cite{wired2024evidence}, which stalled the trial process. Collecting evidence and conducting trials face great challenges. A solution is needed to confirm the authenticity of model outputs and to exclude the possibility of human tampering. 


A similar case happened in finance. News Corp and its subsidiaries, Dow Jones (publisher of The Wall Street Journal) and NYP Holdings (New York Post), are suing Perplexity.AI for copyright infringement. The AI company unlawfully copied and used their articles to train models and generate responses without permission or compensation, thereby diverting revenue from the publishers.

\subsection{Use Case II: FinGPT's Testing Results on Columbia's Copyrighted Exam Questions}



In academia and industry, releasing results without data may limit credibility and trustworthiness. The inference process remains a "black box" to outsiders.  A solution is needed to allow public verification of the results while protecting copyrighted benchmarking questions (i.e., input data).  

For example, we have collected a huge set of Columbia Business School's exam questions and would like to release FinGPT's testing results  without copyright infringement. We need a solution that keep the exam questions private while allowing external readers to verify our testing results.  

\subsection{Use Case III: Protection of Trading Strategies in Academic Contests }


We have been organizing a series of FinRL Contests at academic conferences \cite{wang2025finrlcontestsbenchmarkingdatadriven}, encouraging participants open source their reinforcement learning strategies for stock and cryptocurrency trading tasks. It provides well-designed tasks, abundant financial datasets, near-realistic market environments, and useful starter kits. It has attracted $230+$ participants from 20+ countries and 100+ institutions. The contests also introduce tasks of GPT-engineered signals, which use GPT models to generate and engineer alpha signals for trading.

However, we notice that only 10.6\% of the participants came from the industry. Data privacy and model confidentiality are their main concerns. Participants have concerns of sharing their codes or strategies with contest organizers and the open-source community. To engage more industry participants, we would like to adopt a privacy-preserving evaluation protocol.  This ensures convincing results and also protects participants' codes and trading strategies.

\begin{figure}[t]
    \centering
    \includegraphics[width=0.9\textwidth]{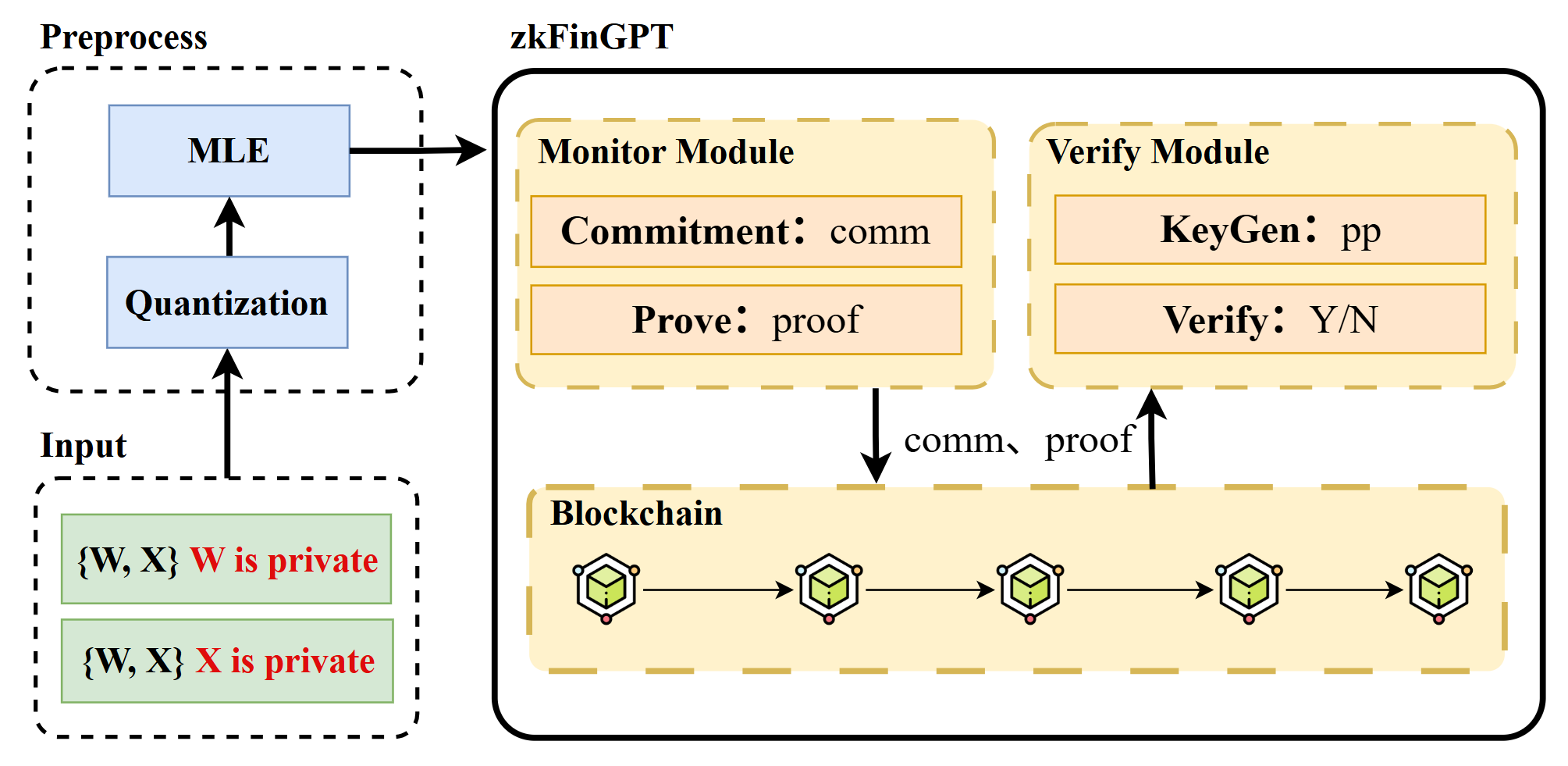}  
    \caption{Pipeline of the proposed zkFinGPT solution.} 
    \label{fig:toolchain}
    \vspace{-0.2in}
\end{figure}

\section{The Proposed zkFinGPT Scheme}

\subsection{Background of Zero-Knowledge Proofs}

We explain the basic idea of the zero-knowledge proof method (ZKP) by considering a simple linear neural network
\begin{equation}
 \label{eq:y=wx}
 \bm{Y} = \bm{W}\bm{X}\in\mathbb{F}^{D_1\times D_3}, ~\text{where}~
 \bm{W}\in\mathbb{F}^{D_1\times D_2},
 \bm{X}\in\mathbb{F}^{D_2\times D_3},
\end{equation}
\begin{equation}
 \label{eq:linear}
 \text{and}~\bm{Y}(i, j) = \sum_{k=0}^{D_2-1} \bm{W}(i, k) \bm{X}(k, j).
\end{equation}
It uses a Multilinear Extension operation (MLE)  and then the sumcheck protocol. 

\subsubsection{Multilinear Extensions}
\label{section:MLE}

The multilinear extension (MLE) \cite{MLE} extends a function defined on a discrete domain (such as a set of matrix indices) into a polynomial function defined continuously over the entire domain and taking consistent values at the same points.

Consider a matrix $\bm{A}\in\mathbb{F}^{D_1\times D_2}$ that defines a discrete function: 
\begin{equation}
\bm{A} : \mathcal{I}_1 \times \mathcal{I}_2 \to \mathbb{F},
\end{equation}
where $\mathcal{I}_1 = \{0, 1, \ldots, D_1-1\}$, $\mathcal{I}_2 = \{0, 1, \ldots, D_2-1\}$. Both indices are using integers.
We rewrite the index as a binary set, i.e.,
\begin{equation}
\bm{A} : \mathcal{B}_1 \times \mathcal{B}_2 \to \mathbb{F},
\end{equation}
where $\mathcal{B}_1 = \{0,1\}^{\log_2 D_1}$ and $\mathcal{B}_2 = \{0,1\}^{\log_2 D_2}$.
The indices for both dimensions are represented by binary strings of length $\log_2 D_1$ or $\log_2 D_2$. If $D_1$ and $D_2$ are not powers of 2, they are padded with zero.
The row index $i\in\{0, D_1-1\}$ and column index $j\in\{0, D_2-1\}$ can be written as
\begin{equation}
\begin{split}
\label{eq:q}
e_i &= (i_1,i_2,\ldots,i_{\log_2 D_1})\in\{0,1\}^{\log_2 D_1}, \\
e_j &= (j_1,j_2,\ldots,j_{\log_2 D_2})\in\{0,1\}^{\log_2 D_2}.
\end{split}
\end{equation}
Therefore, matrix $\bm{A}$ can be written as
\begin{equation}
\bm{A}(e_i,e_j)=\bm{A}(i_1,i_2,\ldots,i_{\log_2 D_1}; j_1,j_2,\ldots,j_{\log_2 D_2})
\end{equation}
The multilinear extension $\widetilde{\bm{A}}(u,v)$ is a multivariate linear polynomial where $u$ and $v$ are continuous variables defined as
follows 
\begin{equation}
u , v\in\mathbb{F}.
\end{equation}
And it satisfies that
\begin{equation}
\label{eq:consistency}
\widetilde{A}(e_i,e_j)=\bm{A}(e_i,e_j),\quad\forall(e_i,e_j)\in\{0,1\}^{\log_2 D_1+\log_2 D_2}.
\end{equation}
MLE distributes each discrete element $\bm{A}(e_i,e_j)$ of the matrix over the continuous domain $\mathbb{R}$. 
When $u$ and $v$ take values in $\{0,1\}^{\log_2D_1}$ and $\{0,1\}^{\log_2D_2}$, then $\widetilde{A}(e_i,e_j)=\bm{A}(e_i,e_j)$, maintaining consistency.

\subsubsection{Schwartz-Zippel Lemma}
\label{section:Schwartz-Zippel}

According to the Schwartz–Zippel lemma \cite{Schwartz} \cite{Zippel}, for a non-zero polynomial $g(x_1,\dots,x_m)$ of degree $d$ over the field $\mathbb{R}$, when we independently select random numbers from a finite field $\mathbb{F} \subseteq \mathbb{R}$, then
\begin{equation}
\text{prob.}[g(r_1,\dots,r_m)=0] \le \frac{d}{|\mathbb{F}|},
\end{equation}
where $|\mathbb F| \gg d$. If the polynomial $g(\cdot)$ is not identically equal to 0, then the probability that a randomly selected point in the field is exactly equal to 0 is low.

In other words, if there are two distinct polynomials of degree $d$, the probability that they are considered functionally equal on the field (i.e., equal at randomly selected points) is extremely small, $\text{prob.} \le d/|\mathbb F|$.

Assume that $g^\prime(\cdot)$ is a forged polynomial, i.e., $g^\prime(\cdot) \neq g(\cdot)$. 
there is a probability of $(1 - d/|\mathbb F|)$ that $g^\prime(\cdot)$ and $g(\cdot)$ are inconsistent. 
$|\mathbb F| \gg d$, so the probability of the verification result being “No” is high.

\subsubsection{Sumcheck Protocol}
\label{section:sumcheck}

The correctness of matrix multiplication (e.g., $\bm{Y}=\bm{W}\bm{X}$) can be verified using the sumcheck protocol \cite{chiesa2017zero} \cite{cormode} on the MLE of these matrices. 

For a polynomial $g(\cdot)$ with $m$ variables over a finite field $\mathbb{F}$, the sumcheck protocol allows the prover to demonstrate the following equation to the verifier,
\begin{equation}
\label{eq:sumcheck}
\sum_{b_1\in\{0,1\}}\sum_{b_2\in\{0,1\}}\dots\sum_{b_m\in\{0,1\}}g(b_1,...,b_m)=C.
\end{equation}
Before the protocol begins, the prover claims that $\sum g = C$, $C$ is a constant. 

\textbf{In the first round.} The prover sends the verifier a univariate polynomial $g_1(x_1)$, claiming that this polynomial satisfies the following conditions, 
\begin{equation}
g_1(x_1)=\sum_{(b_2,\dots,b_m)\in\{0,1\}^{m-1}}g(x_1,b_2,\dots,b_m).
\end{equation}
The verifier checks that the degree of $g_1(x_1)$ is less than or equal to $deg(g_1)$, and that $C = g_1(0) + g_1(1)$. 
The protocol continues if and only if both conditions are met. 
Otherwise, it is rejected. 
If the protocol continues, the verifier selects a random number $r_1 \in \mathbb{F}$ and sends $r_1$ to the prover.

\textbf{In round $\bm{i}$ $(1\le i\le m)$.} The prover sends the verifier a univariate polynomial $g_i(x_i)$, claiming that this polynomial satisfies the following conditions,
\begin{equation}
\label{eq:gi}
g_i(x_i)=\sum_{(b_{i+1},\dots,b_m)\in\{0,1\}^{m-i}}g(r_1,\dots,r_{i-1},x_i,b_{i+1},\dots,b_m).
\end{equation}
The verifier checks that the degree of $g_i(x_i)$ is less than or equal to $deg(g_i)$, and that $g_{i-1}(r_{i-1}) = g_i(0) + g_i(1)$. 
The protocol continues if and only if both conditions are met. 
Otherwise, it is rejected. 
If the protocol continues, the verifier selects a random number $r_i \in \mathbb{F}$ and sends $r_i$ to the prover.

\textbf{In round $\bm{m}$.} The prover sends the verifier a univariate polynomial $g_m(x_m)$, claiming that this polynomial satisfies the following conditions,
\begin{equation}
g_m(x_m)=g(r_1,\dots,r_{m-1},x_m).
\end{equation}
The verifier checks that the degree of $g_m(x_m)$ is less than or equal to $deg(g_m)$, and that $g_{m-1}(r_{m-1}) = g_m(0) + g_m(1)$. 
The protocol continues if and only if both conditions are met. 
Otherwise, it is rejected. 
If the protocol continues, the verifier selects a random number $r_m \in \mathbb{F}$ to verify
\begin{equation}
g_m(r_m) = g(r_1, r_2, \dots, r_m).
\end{equation}
If true, the sumcheck protocol passes, and the verifier believes the prover's claim, $\sum g = C$.

If $g_i(x_i)$ is a pseudo-polynomial given by the prover.
According to the Schwartz–Zippel lemma, 
for two distinct polynomials over a field $\mathbb{F}$, 
the probability that they coincide on a randomly chosen point is at most $\deg(g)/|\mathbb F|$, where $\deg(g)$ is the degree of the polynomial.
The field $\mathbb{F}$ is sufficiently large, sumcheck will fail verification with a very high probability.

The original sumcheck is interactive. 
the Verifier sends a random challenge $r_i$ to the prover in each round and also receives the polynomial $g_i(x_i)$ computed by the prover. 
To achieve non-interactive proofs on the blockchain, we use the Fiat–Shamir transformation \cite{fiat} to generate the random challenge $r_i$. 
The Prover then packages $g_i(x_i)$ and sends it to the verifier all at once, without requiring online interaction.

\subsection{Example Task}


We consider the following simple example of \eqref{eq:y=wx} 
\begin{equation}
\begin{split}
&\bm{W}=\begin{pmatrix} w_{00}=1 & w_{01}=2 \\ w_{10}=3 & w_{11}=4 \end{pmatrix},\quad \bm{X}=\begin{pmatrix} x_{00}=5 & x_{01}=6 \\ x_{10}=7 & x_{11}=8 \end{pmatrix}, \\
&\text{then}~\bm{Y}=\bm{W}\bm{X}=\begin{pmatrix} y_{00}=19 & y_{01}=22 \\ y_{10}=43 & y_{11}=50  \end{pmatrix}, 
\end{split}
\end{equation}
where $\bm{W}$ is prover's proprietary intellectual property, and $\bm{X}$ is publicly model input data.
The inference process is a "black box" for the verifier.
The prover must prove to the verifier that $\bm{Y}$ is indeed the result of the input $\bm{X}$ on the weights $\bm{W}$, without revealing any information about $\bm{W}$.

The sumcheck protocol utilizes the polynomial check of the Schwartz–Zippel in Section \ref{section:Schwartz-Zippel}.
\begin{itemize}[leftmargin=*]

\item If $\bm{Y}$ is the inference output of input $\bm{X}$ on weights W, i.e., $\bm{Y}=\bm{W}\bm{X}$, then the polynomial $g(\cdot)$ is obtained through MLE. After a "random check" by the sumcheck protocol, $\sum g = 0$ is accurate. Since the polynomial $g(\cdot)$ is not identically equal to 0, the probability that it is exactly equal to 0 at a randomly selected point in the field is low.

\item If $\bm{Y}$ is not the inference output of input $\bm{X}$ on weight $\bm{W}$, i.e., $\bm{Y} \neq \bm{W}\bm{X}$, then the polynomial $g^\prime(\cdot)$ is obtained through MLE, $g^\prime(\cdot) \neq g(\cdot)$. The probability of $(1-d/|\mathbb F|)$ to find that $g^\prime(\cdot)$ and $g(\cdot)$ are inconsistent. $|\mathbb F| \gg d$, so the probability of the verification result being "No" is high.

\end{itemize}
Below, we will introduce in detail the specific applications of MLE and the sumcheck protocol.

\subsubsection{MLE in the Example Task}

Each index will use $\log_2 2 = 1$ bit, thus row $i$, index $k$, and column $j$ are Boolean variables, 
\begin{equation}
e_i\in\{0,1\},\quad e_k\in\{0,1\},\quad e_j\in\{0,1\}.
\end{equation}
Therefore, there are a total of $m = 3$ Boolean variables, sumcheck will work over $\{0,1\}^3$.

Define the basis functions for MLE (at boolean point $t\in\{0,1\},~\ell_b(t)=1_{t=b}$),
\begin{equation}
\ell_0(t)=1-t,\qquad \ell_1(t)=t, \quad t\in\mathbb R.
\end{equation}
Therefore, matrix $\bm{W}$, $\bm{X}$ and $\bm{Y}$ can be written as 
\begin{equation}
\begin{split}
\widetilde W(i,k) &= w_{00}\,\ell_0(i)\ell_0(k) + w_{01}\,\ell_0(i)\ell_1(k) + w_{10}\,\ell_1(i)\ell_0(k) + w_{11}\,\ell_1(i)\ell_1(k)  \\
&= 1(1-i)(1-k)+2(1-i)k+3i(1-k)+4ik  \\
&= 1+2i+k,  \\
\widetilde X(k,j) &= x_{00}\,\ell_0(k)\ell_0(j) + x_{01}\,\ell_0(k)\ell_1(j) + x_{10}\,\ell_1(k)\ell_0(j) + x_{11}\,\ell_1(k)\ell_1(j)  \\
&= 5(1-k)(1-j)+6(1-k)j+7k(1-j)+8kj  \\
&= 5+2k+j,  \\
\widetilde Y(i,j) &= y_{00}\,\ell_0(i)\ell_0(j) + y_{01}\,\ell_0(i)\ell_1(j) + y_{10}\,\ell_1(i)\ell_0(j) + y_{11}\,\ell_1(i)\ell_1(j)  \\
&= 19(1-i)(1-j)+22(1-i)j+43i(1-j)+50ij  \\
&= 19+24i+3j+4ij.
\end{split}
\end{equation}
In order to convert the matrix multiplication equation into the “sum of the whole equals 0” form used in sumcheck, we define a polynomial
\begin{equation}
\begin{split}
g(i,k,j)&= \frac{1}{2}\widetilde Y(i,j)-\widetilde W(i,k)\,\widetilde X(k,j)  \\
&= 4.5+2i+0.5j-9k-4ik-kj,
\end{split}
\end{equation}
the variable names still use $i$, $j$, $k$, but now $i, j, k \in \mathbb{G}$. 
Since $\widetilde W$, $\widetilde X$ and $\widetilde Y$ are obtained through MLE, $g(\cdot)$ is a polynomial with a degree of no more than 1 in each variable.

When $i,j,k\in\{0,1\}$, $\widetilde W$, $\widetilde X$ and $\widetilde Y$ recover to the original matrix values, e.g. $\widetilde W(0,1)=w_{01}$, therefore
\begin{equation}
\sum_{k\in\{0,1\}} g(i,k,j) = \frac{1}{2}Y(i,j) - \sum_{k} W(i,k)X(k,j),
\end{equation}
the statement $\bm{Y}=\bm{W}\bm{X}$ is equivalent to $\sum g = 0$.

To combine the checks for all $i$, $j$ into a single sumcheck task, it is simpler to directly take $i, k, j$ as variables for the sumcheck.
We order the three variables as $b_1 = i$, $b_2 = k$, $b_3 = j$,
\begin{equation}
\sum_{b_1,b_2,b_3\in\{0,1\}} g(b_1,b_2,b_3) = 0,
\end{equation}
therefore, $g(\cdot)$ is a three-variable polynomial.

\subsubsection{Sumcheck Protocol in the Example Task}

This is the target polynomial that the prover needs to prove using sumcheck,
\begin{equation}
g(b_1, b_2, b_3) = 4.5+2b_1+0.5b_3-9b_2-4b_1b_2-b_2b_3.
\end{equation}
Before the sumcheck protocol begins, the prover claims that $\sum g = C = 0$. 

\textbf{In the first round}, the prover computes $g_1(X_1)$,
\begin{equation}
g_1(x_1) = g(x_1, 0, 0) + g(x_1, 0, 1) + g(x_1, 1, 0) + g(x_1, 1, 1),
\end{equation}
i.e., $g_1(x_1) = 0$. 
The prover sends $g_1(x_1)$ and $C = 0$ to the verifier.
To prevent tampering, the files are uploaded to the blockchain.
The verifier checks that the degree of $g_1(x_1)$ is at most $deg(g_1) = 1$, then verifies that $g_1(0) + g_1(1) = C = 0$. 
Then sends a random number $r_1 = 2$ to the prover.

\textbf{In the second round}, the prover computes $g_2(x_2)$,
\begin{equation}
g_2(x_2) = g(r_1, x_2, 0) + g(r_1, x_2, 1),
\end{equation}
i.e., $g_2(X_2) = 17.5-35b_2$. 
The prover sends $g_2(x_2)$ and $g_1(r_1) = 0$ to the verifier.
To prevent tampering, the files are uploaded to the blockchain.
The verifier checks that the degree of $g_2(x_2)$ is at most $deg(g_2) = 1$, and verifies that $g_2(0) + g_2(1) = g_1(r_1) = 0$. 
Then sends a random number $r_2 = 3$ to the prover.

\textbf{In the third round}, the prover computes $g_3 (x_3)$,
\begin{equation}
g_3(x_3)=g(r_1, r_2, x_3),
\end{equation}
i.e., $g_3(x_3) = -42.5-2.5b_3$.
The prover sends $g_3(x_3)$ and $g_2(r_2) = -87.5$ to the verifier.
To prevent tampering, the files are uploaded to the blockchain.
The verifier checks that the degree of $g_3(x_3)$ is at most $deg(g_3) = 1$, and verifies that $g_3(0) + g_3(1) = g_2(r_2) = -87.5$.
Then take the random number $r_3=4$ to get $g(r_1, r_2, r_3) = g(2, 3, 4) = -52.5$, and computes $g_3(r_3) = -52.5$. 
The following equation is correct 
\begin{equation}
g_3(r_3)=g(r_1, r_2, r_3),
\end{equation}
and verification passed.

In conclusion, the sumcheck protocol passes, and the verifier believes the prover's claim, $\sum g = 0$.

\subsection{Overview of zkFinGPT Scheme} 

Consider an LLM with weights $\bm{W}$, the input is $\bm{X}$, and the output is $\bm{Y}$, as shown in Fig. \ref{fig:toolchain}.
Note that $\bm{W}$ is the intellectual property of the prover, while $\bm{X}$ is the public input data.
The inference process is a "black box" for the verifier.
The prover must prove to the verifier that $\bm{Y}$ is indeed the result of the input $\bm{X}$ on the weights $\bm{W}$, without revealing any information about $\bm{W}$.

To address this, we propose zkFinGPT that uses ZKP \cite{zkp} to make FinGPT's inference process publicly verifiable while preserving privacy. 
It verifies model legitimacy and output credibility.
Not only that, in order to make sure that the commitment file and proof file are immutable, we upload them onto the blockchain to ensure that they cannot be deleted.

As shown in Fig. \ref{fig:toolchain}, zkFinGPT has two types of inputs: "$\bm{W}$ is private", and "$\bm{X}$ is private".
We input $\bm{W}$ and $\bm{X}$, where $\bm{W}$ is private and $\bm{X}$ is public.
Firstly, zkFinGPT quantizes $\bm{W}$, $\bm{X}$ into integers over a finite field $\mathbb{F}$. 
Then converts them into polynomials usable with zkSNARK \cite{wahby2018doubly} using MLE.
Finally input them into zkFinGPT.

The zkFinGPT consists of two modules: \textsf{Monitor} (Commit \& Prove) and \textsf{Verify} (KeyGen \& Verify). For security, the \textsf{Verify} module is deployed in the cloud. To ensure the commitment and proof files are immutable and cannot be deleted, uploads them to the blockchain.

\begin{itemize}[leftmargin=*]
    \item \textsf{Monitor} module tracks the inference process and generates commitment and proof files. These files will be uploaded to the blockchain.

        \begin{itemize}[leftmargin=*]

        
        \item zkFinGPT.Commit($pk$, $\bm{W}$): Run $comm$ $\leftarrow$ zkFinGPT.Commit($pk$, $\bm{W}$) and sends verifier the commitment file $comm$. 
        As shown in \eqref{eq:comm}, when $p_i$ is model weights, $pk$ is used to convert the model weights $\bm{W}$ into a commitment point on the elliptic curve. The computational overhead is extremely high.

        \item zkFinGPT.Prove($pk$, $\bm{W}$, $u$): Run ($v$, $\pi$) $\leftarrow$ zkFinGPT.Prove($pk$, $\bm{W}$, $u$) and outputs $v$ and proof file $\pi$. 
        KZG \cite{kzg} open commitment at point $u$ and submits opening to the verifier (see Appendix \ref{section:kzg} Opening Creation).
        $\pi$ includes a sequence of random challenges, the opening, and the univariate polynomials $g_i(x_i)$ \eqref{eq:gi} of sumcheck protocol computed by the prover in each round.

        \end{itemize}
        
    \item \textsf{Verify} module generates keys and verifications. The $comm$ and $\pi$ will be fetched from the blockchain. 
    
        \begin{itemize}[leftmargin=*]

        \item zkFinGPT.KeyGen($1^\lambda$, $r$): Run $(pk, vk)$ $\leftarrow$ zkFinGPT.KeyGen($1^\lambda$, $r$) and outputs $pk$ (prove key) and $vk$ (verify key) \eqref{eq:key}. $\lambda$ is a security parameter, usually 128.
        The size of the finite field $|\mathbb{F}| \approx 2^\lambda$.
        $r$ is a random number. 

        \item zkFinGPT.Verify ($vk$, $comm$, $\pi$, $u$, $v$): Run Y/N $\leftarrow$ zkFinGPT.Verify ($vk$, $comm$, $\pi$, $u$, $v$) and outputs Yes or No. 
        The verifier will use the sumcheck protocol to verify the correctness of $\bm{Y}=\bm{W}\bm{X}$ and the KZG to verify the consistency of $\bm{W}$ (see Appendix \ref{section:kzg} Opening Verification).
        
        \end{itemize}
        
\end{itemize}

\vspace{-0.1in}
\subsection{zkFinGPT for Case I}

\begin{figure}
    \centering
    \includegraphics[width=0.9\textwidth]{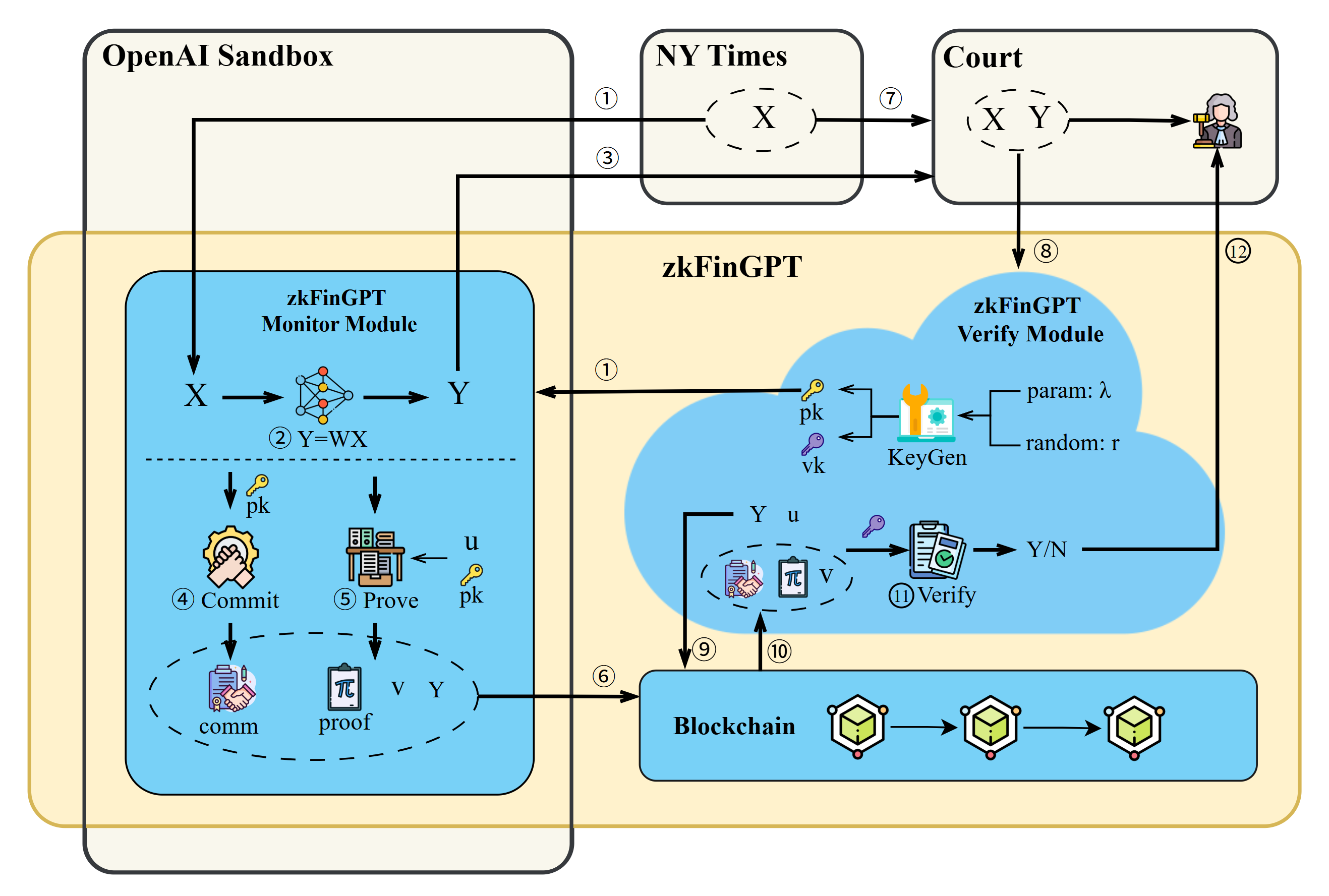}  
    \caption{Overview of zkFinGPT solution for Case I.} 
    \label{fig:case_one_zkp}
    \vspace{-0.2in}
\end{figure}

We design zkFinGPT as third-party software to verify the inference log file and protect the model weights $\bm{W}$, as shown in Fig. \ref{fig:toolchain}.  zkFinGPT enables courts to confirm the authenticity of model output, as shown in Fig. \ref{fig:case_one_zkp}.

First, NYT collects evidence, where the prompt $\bm{X}$ is the input and $\bm{Y}_0$ is the original allegedly plagiarized NYT article.
In this sandbox, OpenAI performs inference, commit, and prove. The entire process was completed under the zkFinGPT \textsf{Monitor} module.
The records $(\bm{X}, \bm{Y})$, $comm$, $\pi$ are uploaded onto a blockchain that is immutable. The new evidence $(\bm{X}, \bm{Y}, \bm{Y}_0)$ is submitted to the court. During the trial, the court sends $(\bm{X}, \bm{Y})$ to the zkFinGPT \textsf{Verify} module.  

The zkFinGPT \textsf{Verify} module uses $vk$, $\bm{X}$, and $comm$ and $\pi$ fetched from the blockchain to check the trustworthiness of $\bm{Y}$. The verification result supports the court to judge whether OpenAI is engaging in fraud. 
By the zero-knowledge property of ZKP, neither NYT nor the court can learn any information about $\bm{W}$ from $comm$ and $\pi$.  At the same time, blockchain technology, through its decentralized mechanism and encrypted chain storage structure, achieves tamper-proof and undeletable log files. By adopting the zkFinGPT, there is no need to worry about OpenAI engineers deleting  records. This makes the trial smoother and more reliable.

\subsection{zkFinGPT for Case II}

In Case II, let $\bm{Y}=\bm{W}\bm{X}$, where $\bm{W}$ is the FinGPT model weights, $\bm{X}$ is the copyright exam set, and $\bm{Y}$ is the inference result. As shown in Fig. \ref{fig:toolchain}, zkFinGPT can prove that the output is inferred by the claimed model and dataset without disclosing sensitive input data. It makes the published inference results more convincing. The publisher of the inference result commits to and proves to the inference process, obtaining $comm$ and $\pi$. 
These log files can be used by zkFinGPT to verify the trustworthiness of $\bm{Y}$. The verifier decides whether to accept $\bm{Y}$ based on the verification result. 


\subsection{zkFinGPT for Case III}

In Case III, the testing process is abstracted as $\bm{Y}=\bm{W}\bm{X}$, where $\bm{W}$ is the submitted trading strategy, $\bm{X}$ is the data in the given trading task, and $\bm{Y}$ is the trading agent's actions.  We adopt zkFinGPT to allow external investors to verify the trading actions while protecting participant's trading strategy, as shown in Fig. \ref{fig:toolchain}. FinRL contest commits to and proves to the testing process, obtaining $comm$ and $\pi$. These log files allow zkFinGPT to verify the trustworthiness of $\bm{Y}$. Investors decide whether to accept $\bm{Y}$ based on the verification results. 


\vspace{-0.1in}
\section{Experimental Results for Computational Overhead}

\begin{table}[t]
  \caption{Computational overhead of the Llama models using the zkLLM package \cite{sun2024zkllm}.}
  \label{overhead-table}
  \centering
  \begin{tabular}{lcccccccc}
    \toprule
    Model     & \makecell{Layer\\num} & \makecell{Commit\\time (s)} & \makecell{Proof\\size (MB)} & \makecell{Prove\\time (s)} & \makecell{Verify\\time (s)}\\
    \midrule
    Llama2-13B & 40 & 986 & 11.0 & 803 & 3.95  \\ 
    Llama3-1B & 16 & 76 & 1.33 & 433 & 2.26  \\ 
    Llama3-8B & 32 & 531 & 7.97 & 620 & 2.36  \\
    Llama3-70B & 80 & 5310 & 25.35 & 1578 & 4.66  \\ 
    \bottomrule
  \end{tabular}
\end{table}

\begin{table}[t]
  \caption{Computational overhead of the GPT2 models using the zkGPT package \cite{qu2025zkgpt}.}
  \label{overhead-table}
  \centering
  \begin{tabular}{lcccccccc}
    \toprule
    Model     & \makecell{Layer\\num} & \makecell{Commit\\time (s)} & \makecell{Proof\\size (KB)} & \makecell{Prove\\time (s)} & \makecell{Verify\\time (s)}\\
    \midrule
    GPT2-0.1B & 12 & 44.112 & 88.344 & 33.799 & 0.422  \\ 
    \bottomrule
  \end{tabular}
  \vspace{-0.2in}
\end{table}

\subsection{Experiment Settings}

\textbf{Software.} Our initial experiments used the packages zkLLM \cite{sun2024zkllm} and zkGPT \cite{qu2025zkgpt}, implemented in CUDA C++. We use the sumcheck protocol \cite{chiesa2017zero} \cite{cormode} and the KZG polynomial commitment scheme \cite{kzg}.

\textbf{Hardware.} 
Our testings were conducted on a server with 112-core Intel(R) Xeon(R) Gold 6330 CPU (2.00 GHz, 1.5TB Memory) and an NVIDIA A100 (80 GB).

\textbf{Models and Datasets.} Models include LLama and FinGPT. Datasets include copyrighted exam questions and financial data.

\subsection{Computational Overhead}

The application of ZKP to large GPT models incurs high computational costs.  Table \ref{overhead-table} shows zkLLM's overhead on LLama models. 
For a GPT model with $N$ layers and size $M$ (in billions), the results indicate the proving time complexity is $O(N)$, from 620s for 32 layers (7B) to 1578s for 80 layers (70B). 
The verification time complexity is $O(\sqrt{N})$, which remains efficient, below 5s for all models.
Commitment time complexity is $O(M)$, from 531s (7B) to 5310s (70B), with commitment sizes of 7.97 to 25.35MB.

From the results, the proving time is typically hundreds of times longer than the model inference, and the commitment time is also significant.
Since the zero-knowledge proof process requires intensive computation, generating a single proof is far more computationally expensive than inference.
Therefore, computational overhead has become the primary bottleneck in applying zkFinGPT, making performance optimization crucial.

The current zkLLM uses 16-bit quantization, leaving room for optimization. We are focusing on reducing quantization precision and scaling to larger GPT models to reduce proof time and commitment size.

\subsection{Profiling}

\section{Conclusion and Future Work}

This paper proposed a novel zkFinGPT solution that used ZKP. This solution can verify the credibility of a GPT's inference process without compromising the privacy of model weights or input data. It has promising values in three financial use cases. For future research, we will apply quantization and GPU parallel computing techniques to improve zkFinGPT performance and reduce proof time and file size for large GPT models.

\bibliographystyle{unsrt}
\bibliography{ref}

\appendix

\section{KZG Commitment Scheme}
\label{section:kzg}

The KZG polynomial commitment scheme \cite{kzg} guaranties the consistency of privacy (such as $\bm{W}$) during the verification of the correct computation. This scheme is based on elliptic curve cryptography. It allows a prover to commit to a polynomial. Then expose the polynomial to a verifier at a random point for verification while keeping the polynomial hidden from the verifier and all other parties.

Suppose an n-degree polynomial $P(x)$:
\begin{equation}
P(x)=\sum_{i=0}^{n}p_ix^i, 
\end{equation}
where $p_i$ are the polynomial coefficients, $p_i$ and $x$ are elements of the finite field $\mathbb{F}$. 
KZG are based on pairwise friendly elliptic curves, such as BLS12–381 \cite{bls}, which are defined over $\mathbb{F}$, contain two elliptic curve groups $G_1$ and $G_2$, and support pairwise $e$.

KZG consists of four phases: trusted setup, commitment, opening creation, and opening verification.
\begin{itemize}[leftmargin=*]

  \item \textbf{Trusted Setup.}   
  Unlike traditional commitment schemes, KZG requires a trusted setup.
  First, a secret key $s$ is generated, which is a random number from $\mathbb{F}$.
  $G$ from the elliptic curve group $G_1$ and is the generator used for commitment. 
  $H$ from the elliptic curve group $G_2$ and is the generator used for verification. 
  As shown in \eqref{eq:key}, $G$ and $H$ are each multiplied by $n$ random scalars, forming $pk$ (prove key) and $vk$ (verify key) respectively, used for proof and verification. 
  These guarantee that each commitment point is a random and independent base point, which cannot be forged by the prover.
  \begin{equation}
  \begin{split}
    \label{eq:key}
    pk&=s^iG=(G,sG,s^2G,\ldots,s^nG), \\
    vk&=(H,sH,s^2H,\ldots,s^nH).
  \end{split}
  \end{equation}

  \item \textbf{Commitment.} 
  The prover can use $pk$ to generate commitment ($comm$) to the polynomial $P(x)$: 
  \begin{equation}
    \label{eq:comm}
    \mathrm{comm}=\mathrm{Commit}(P)=P(s)G=\sum_{i=0}^np_is^iG.
  \end{equation}

  \item \textbf{Opening Creation.} 
  The prover must open the polynomial to the verifier at a random point (from verifier). i.e., provide the verifier with an evaluation proof (opening) at a random point.
  Suppose the prover wants to prove to the verifier that $P(x)$ evaluates to the value $v$ at point $u$, i.e., $P(u) = v$. We can obtain:
  \begin{equation}
    \large Q(x)=\frac{P(x)-v}{x-u}.
  \end{equation}
  In fact, $Q(x)$ is a polynomial of $(n-1)$ degree. 
  In a manner similar to commitment $comm$, the opening (i.e., the proof that evaluates to $P(u)=v$) is generated as follows: 
  \begin{equation}
    \mathrm{opening}=Q(s)G=\sum_{i=0}^{n-1}q_is^iG,
  \end{equation}
  where $q_i$ are the coefficients of the quotient polynomial $q(x)$.

  \item \textbf{Opening Verification.} 
  Given commitment, evaluation, and corresponding proof (opening), verifier can verify the opening using pairwise operations. The $comm$ and opening are valid if the following equation holds,
  \begin{equation}
  \label{eq:openingVerify}
    e(comm,H)=e(opening,sH/uH)e(G,H)^{v}.
  \end{equation}
  
\end{itemize}

\end{document}